\documentclass[preprint,tightenlines]{revtex4}
\usepackage{graphicx}
\usepackage{bm}

\begin{document}
\title{\bf Many-body luminescence from highly excited quantum-confined
  structures}
\author{T. V. Shahbazyan}
\affiliation{Department of Physics, Jackson State University, Jackson,
  MS 39217}
\author{M. E. Raikh}
\affiliation{Department of Physics, University of Utah, Salt Lake City,
 UT 84112}
 \begin{abstract}
We review recent results on many-body effects in the luminescence from
semiconductor nanostructures.  Many-body luminescence from highly
excited quantum-confined structures is {\em conceptually} important
topic since a new parameter, a level spacing, plays a crucial
role. This spacing is {\em not} merely a discretization of the bulk
luminescence spectrum, as it could seem. The {\em interplay} of finite
spacing with interactions (even weak) results in a highly nontrivial
sequence of emission lines, their heights revealing the many-body
correlations in the system. Here the complex structure of the emission
spectrum, resulting from the shakeup processes in many-particle (but
finite) system, is demonstrated for a confined electron-hole system of
a particular geometry, in which the interacting carriers are
confined to a ring. For this geometry, the Luttinger liquid theory
allows one to exactly calculate the {\em intensities} of all many-body
spectral lines. The {\em positions} of the lines are governed by the
relation of the level spacings for electrons and holes.  While close
to the emission threshold the interactions cause only weak shakeup
satellites of the single-particle lines, away from the threshold the
discrete luminescence spectrum is completely dominated by the
many-body transitions.  We describe the Luttinger liquid approach for
calculations of optical spectra in finite one-dimensional systems.
The calculations are preceded by a detailed review of experimental
and theoretical work on many-body luminescence from various infinite
systems.  We also review the current status of the experimental and
theoretical research on quantum nanorings.
\end{abstract}
\maketitle

\section{INTRODUCTION}

Luminescence from zero-dimensional objects (quantum dots)
is one of the highlights in physics of nanostructures which emerged
during the last decade. Early papers (see, e.g.,
Refs.\ \cite{brunner92,marzin94}, and the review article Ref. \cite{zrenner00})
reported the PL spectra consisting of "zero-width" luminescence lines.
High surface density of quantum dots caused an ambiguity in assigning
of these lines. In the later studies the emission from a {\em single} dot was
resolved. This progress\cite{gammon00} has permitted the luminescence
spectroscopy of individual dots with controllable exciton  population
determined  by the excitation intensity.

Since many-body optical transitions in zero-dimensional objects were
demonstrated experimentally, it is important to assess this
phenomenon from the  perspective of  the well established field
of many-body luminescence. This is accomplished in the present
chapter. Below we  review the many-body luminescence in various
systems studied to date experimentally and theoretically. We then
demonstrate that many-body luminescence from highly excited
zero-dimensional objects has unique features due to {\em large} number
of {\em discrete} lines. This discreteness {\em unravels} the many-body
correlations that are otherwise masked in the continuous spectrum of
luminescence from infinite systems. We describe in detail the
emergence of such correlations for a particular nanostructure geometry
-- semiconductor nanorings -- using the Luttinger liquid approach for
quasi-one-dimensional finite-size systems.

\section{SPECTROSCOPY OF MANY-BODY PROCESSES}

\subsection{Shakeup effects in optical spectra of many-electron systems}

Shakeup represents a fundamental many-body effect that takes place in optical
transitions in many-electron systems. In such systems, an absorption or
emission of light is accompanied by electronic excitations in the final state
of the transition.  The most notable shakeup effect is the Anderson
orthogonality catastrophe\cite{anderson67} in the electron gas when the initial
and final states of the transition have very small overlap due to the
readjustment of the Fermi sea electrons in order to screen the Coulomb
potential of photoexcited core hole. Shakeup is especially efficient when the
optical hole is immobilized, and therefore it was widely studied in
conjunction with the Fermi edge singularity (FES) in metals
\cite{mahan67,nozieres69,langreth70} and doped semiconductor quantum wells
\cite{chemla87,ruckenstein86,ruckenstein87,skolnik87,chemla88,chemla89,kalt89}.
Comprehensive reviews of FES and related issues can be found in
Refs. \cite{mahan,ohtaka90}.

\subsubsection{Shakeup processes in electron gas}

The long-time dynamical Fermi sea response to a sudden appearance of
the optical hole Coulomb potential can be viewed as a dressing of that
hole by the low-energy Fermi sea excitations. This leads to the power-law
infrared divergence in the hole density of
states\cite{mahan,ohtaka90}. In three-dimensional (3D) electron gas,
the only low-energy excitations are the Fermi sea electron-hole pairs,
and, therefore, close to the absorption onset, the electron-electron
interactions are usually neglected. Incorporation of electron-electron
interactions gives rise to the plasmon satellites
\cite{langreth70,livins88}, which are somewhat similar to the
low-energy phonon replicas.  Each of these satellites also represents
a power-law divergence at the energy $n\omega_p$, where $\omega_p$ is
the bulk plasmon frequency (we set $\hbar=1$) and $n$ is an
integer. In contrast, in low-dimensional electron systems, the plasmon
is gapless, so the shakeup of single-particle and collective
excitations must be treated on equal footing. In 2D electron gas,
where the plasmon dispersion is $\omega_q\propto \sqrt{q}$ for small
$q$, the plasmon shakeup leads to a narrowing of the main singularity
and to an additional structure at energies corresponding to the
plasmon bandwidth \cite{hawrylak90}. Note, however, that plasmon
effect is negligible near the Fermi edge because of a much smaller 2D
plasmon density of states at low energies as compared to that of
electron-hole pairs. In contrast, the role of plasmon shakeup is much
more important in 1D electron systems, where the plasmon dispersion is
linear. Since the electron dispersion in the vicinity of the Fermi
level is also linear, single-particle and collective excitations are
intertwined, forming a strongly correlated electron liquid. The exact
solution of the 1D FES in the long time limit, which was carried out
using the Luttinger model\cite{haldane81}, revealed that the power-law
exponent is determined by both electron-hole and electron-electron
interactions \cite{gogolin93,prokof'ev94,kane94}.

\subsubsection{Magnetoplasmon shakeup in semiconductor
  quantum wells}

Shakeup processes are quite pronounced in luminescence from a 2D electron gas
in a perpendicular magnetic field. In such systems, the single-particle energy
spectrum represents a staircase of equidistant Landau levels (LLs) separated
by the cyclotron energy $\omega_c$. Here, a recombination of an interband
electron-hole pair is accompanied by electronic transitions across the
cyclotron gap. Shakeup satellites, corresponding to excitation of
magnetoplasmons as well as to inter-LL Auger transitions, were observed in a
number of experiments
\cite{potemski91,butov92-shakeup,nash93,skolnick93,skolnick94,finkelstein97,scholz97,manfra98,ando02},
in agreement with earlier theoretical predictions
\cite{sham89,ando89,hawrylak91}. The general expression for the ground state
luminescence intensity has the form\cite{mahan}
\begin{equation}
\label{general}
I(\omega)\propto\sum_{f} C_{f}\, \delta (\omega+E_f - E_i),
\end{equation}
where $E_i$ and $E_f$ are initial (ground) and final state energies and the
oscillator strengths $C_{f}$ are given by the square of the dipole matrix
element. In the absence of interactions, the recombination act does not
perturb the system and so the initial and final state energy difference is simply
$E_g+(\omega_c^e+\omega_c^h)/2$, where both electron and hole belong to the
lowest LL\cite{sham87}. In the presence of interaction, the inter-LL
magnetoplasmon satellites appear in the lower tail of the spectrum at
frequencies that are multiples of the electron cyclotron energy,
$\omega_c^e$. Note that in doped quantum wells, the actual separation
between LLs is less than $\omega_c^e$ due to the exchange effects, so the
magnetoplasmon energy lies above that of single-particle transitions.
Magnetic-filed dependence of the oscillator strengths $C_f$ is determined by
several factors. For even integer filling factors, $\nu=2\pi l^2 n_e$ ($n_e$
is electron concentration and $l$ is the magnetic length), the screening of
Coulomb interaction by the electron gas is strongly suppressed, resulting in
the enhancement of satellite amplitudes\cite{sham89,scholz97}. Another
widely-observed feature was a suppression of satellite peaks for filling factor
$\nu < 2$, i.e., when only the lowest LL is occupied. Such a suppression
originates from the electron-hole symmetry in the lowest LL\cite{lerner81},
which results in a cancellation of the electron and hole final state Coulomb
matrix element contributions to $C_f$ \cite{sham87}. In the valence band,
similar inter-LL shakeup processes were observed in $p$-doped
\cite{glazberg01,kubisa03} GaAs quantum wells.

\subsubsection{Spin wave shakeup in quantum Hall ferromagnets}

\begin{figure}
\centering
\includegraphics[width=4in]{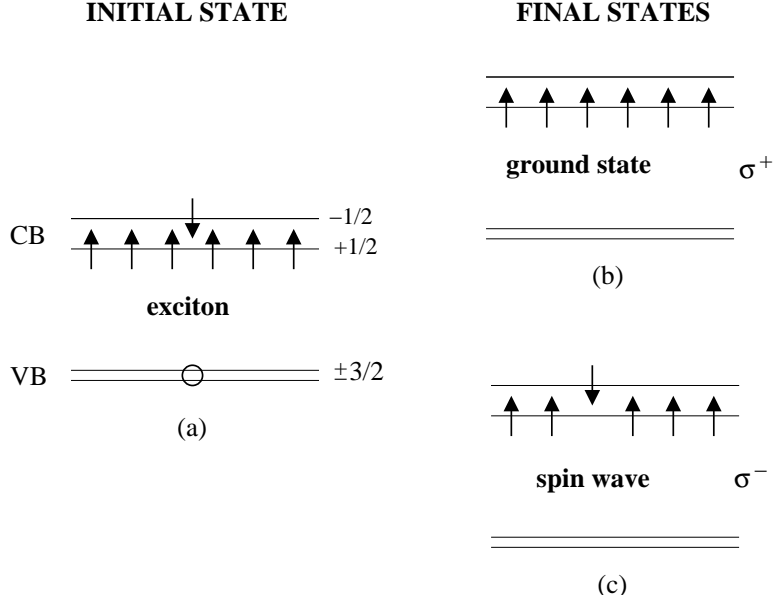}
\caption{\label{fig:spin-wave} Luminescence from $n=0$ LL at filling
  factor $\nu=1$. Recombination of initial interband exciton (a) can
  result either in ground state for $\sigma^{+}$ polarization (b), or
  spin wave shake up for $\sigma^{-}$ polarization (c).
 }
\end{figure}

In one-side modulation-doped quantum wells, the aforementioned electron-hole
symmetry is violated due to the spatial separation between the electron and
valence hole planes that is caused by the interface potential. In such
samples, polarization-dependent spectral redshifts were reported as the
filling factor was swept through integer values
\cite{skolnick94,finkelstein97,manfra98,gravier98,osborn98,takeyama99,munteanu00}.
These redshifts were attributed to the competition between interband
Coulomb binding and electron self-energies in the final
state\cite{cooper97,hawrylak97}. For example, for $\nu=1^-$, the
initial state represents a valence hole and a full spin-polarized LL
in conduction band which is negatively charged in order to compensate
the positive hole charge. A recombination leaves a hole in the
conduction band LL with exchange energy $E_{ex}=\sqrt{\pi/2}\,
e^2/\kappa l$ ($\kappa$ is the dielectric constant). For $\nu=1^+$,
the initial state consists of an interband magnetoexciton made up
from a valence hole and an electron in the upper (spin-up) polarized
LL, with a binding energy $E_0$. Then the removal of an electron from
the lower (spin-down) polarized LL leaves a {\em spin wave} in the
final state of the transition with just the Zeeman energy (see
Fig. \ref{fig:spin-wave}). The difference between below and above $\nu=1$
final state energies is thus $E_{ex}-E_0>0$ in such asymmetric
structures, which accounts for the redshift. A similar analysis was
applied to the redshifts observed near $\nu=2$ filling factor
\cite{skolnick94,finkelstein97,manfra98,gravier98,munteanu00}. In
single heterojunctions, systematic studies of the electron-hole
separation as a function of carrier concentration were carried out in
Ref. \cite{kim01}.

The shakeup of spin waves, mentioned above, has been observed
prominently in the magnetoluminescence spectra at filling factors
close to 1, 3, and 5
\cite{plentz98,manfra98,gravier98,osborn98,takeyama99,munteanu00}.  At
$\nu=1$, an asymmetric broadening of the luminescence peak with
additional spectral weight on low energy side was observed for
$\sigma_{-}$-polarized light \cite{plentz98}. The origin of this
broadening was explained as follows\cite{cooper97}.  For $\sigma_{+}$
polarization, the recombination of an exciton, made up from a valence
hole and an $n=1$ LL electron, leaves the 2DEG in the ground state,
resulting in narrow emission peak (see Fig. \ref{fig:spin-wave}).
For $\sigma_{-}$ polarization, the final state is a spin-wave with
momentum $q$ equal to that of the exciton before the recombination, so
the lineshape at low energies is determined by the difference between
the dispersions of a spin wave and an exciton. Furthermore, for filling
factors $\nu=3$ and 5, a splitting of the ground state luminescence
peak was observed for $\sigma_{-}$
polarization\cite{gravier98,manfra98}. This splitting,
greatly enhanced at low temperatures, was attributed to the
Anderson-Fano resonance in the final state that originated from the
destructive interference between the hole, left in the conduction band
LL after recombination, and the continuous spectrum of the spin wave
\cite{hawrylak97}.

\subsection{Shakeup of electron excitations in few-particle systems}

\subsubsection{Shakeup satellites in atoms}

In atoms and molecules, shakeup satellites, corresponding to internal
electronic transitions, are routinely observed using photoelectron and resonant
Raman spectroscopy. In particular, shakeup satellites can be observed in the
two particle spectrum, i.e., when two holes are left in the final state of an
atom after electron emission. Satellite's strength can be strongly enhanced in
the presence of a resonant intermediate state. For example, in copper atoms,
the incident photon can first excite the core 3p electron to the 4s shell; the
core hole then decays to the 3d shell through the Auger process (with electron
ejected from 3d shell) leaving two 3d holes in the final
state\cite{zangwill81}. For recent reviews of extensive literature the reader
is referred to Refs.  \cite{gel'mukhanov99,armen00}).

\subsubsection{Many-body luminescence lines in single quantum dots}

In semiconductors, multiple emission lines have been recently observed in the
luminescence from single self-assembled quantum dots
\cite{ikezawa97,landin98,bayer98,kulakovskii99,dekel98,dekel00a,dekel00b,findeis00,bayer00,dekel01,regelman01-prl,regelman01-prb,forchel01-prb,forchel02-prb}.
At low excitation intensities, the luminescence from a dot is dominated by the
ground state single-exciton recombination (i.e., from lowest
size-quantization levels
in conduction and valence bands). Upon increasing the excitation intensity, as
the number of excited carriers increases, new emission lines appear in the
luminescence spectrum. These lines were interpreted in terms of
recombination within
many-exciton complexes.  One reason for the emergence of additional lines is
that the carriers, constituting a complex, occupy higher size-quantization
levels. Another reason, is that the interactions between strongly confined
photoexcited carriers lift the {\em degeneracies} within the final many-body
states. The latter mechanism was addressed in Refs.
\cite{dekel98,dekel00a,forchel01-prb,forchel02-prb,hawrylak99} for the
situations with\cite{forchel01-prb,forchel02-prb,hawrylak99} and
without\cite{dekel98,dekel00a} orbital degeneracy of single-particle states.
The calculations carried out in
Refs. \cite{dekel98,dekel00a,forchel02-prb,hawrylak99} predicted the
splittings of luminescence lines, originating from different many-body final
states, to be of the order of matrix element of the interaction
potential. The actual positions of these lines, corresponding to interband
transitions between states with the {\em same} size-quantization quantum
numbers, reproduce quite accurately the experimental PL spectra of
Refs.\ \cite{dekel98,dekel00a,dekel00b,dekel01} (for up to $N=16$) and of
Ref.\cite{bayer00,forchel02-prb} (for $1\leq N \leq 6$).  The third type
of additional lines were identified with the transition energies
corresponding to {\em different} size-quantization
levels\cite{dekel98,dekel00a}. Such transition energies point to
shakeup processes in a {\em confined electron-hole system},
when the recombination of an electron-hole pair is accompanied by internal
excitations within the exciton multiplex.

\subsubsection{Density dependence of optical spectra}


The evolution of optical spectra with increasing electron gas density was
studied both in metals \cite{combescot71} and in semiconductor quantum wells
\cite{hawrylak91-X-FES,brown96,huard00,yusa00}. In the absence of conduction
electrons, the absorption spectrum is characterized by two peaks -- a sharp
exciton line well separated from a step-like onset of continuum states. In
metals, the presence of conduction band electrons asymmetrically broadens the
exciton peak according to $(\omega - \omega_1)^{-\alpha}$ with $\alpha$
decreasing in the (0,1) interval as the electron concentration increases,
while the step-like continuum edge also acquires a power-law shape, $(\omega -
\omega_2)^{\beta}$ with $\beta$ increasing with concentration in the (0,3)
interval \cite{combescot71}.  In contrast, in semiconductor quantum wells with
low electron concentration, the ground state represents a negatively charged
exciton, $X^{-}$, formed as a result of the binding of an additional
conduction electron by a photoexcited interband electron-hole pair
\cite{kheng93,finkelstein95,shields95}. This leads to an emergence of the
exciton peak ($X$), corresponding to ionized $X^{-}$, located between the
ground state $X^{-}$ peak and the absorption onset at the Fermi energy $E_F$
\cite{hawrylak91-X-FES}.  The evolution of absorption/emission spectra with
increasing 2D electron gas concentration was traced in
Refs. \cite{brown96,huard00,yusa00}. As the electron concentration increases,
so does the separation between $X^{-}$ and $X$ peaks; at the same time, the
oscillator strength of $X$ is reduced as the exciton binding energy becomes
smaller than $E_F$. The lineshape undergoes a qualitative change from
exciton-like to continuum-like as concentration exceeds some characteristic
value\cite{brown96,huard00,yusa00}.  In particular, above that value, the
absorption spectrum develops higher-energy tail corresponding to the shakeup
transitions. It was also noted \cite{yusa00} that, in a weak magnetic field,
the character of shakeup satellites in the low-energy tail of luminescence
spectrum changes: for higher electron concentrations, the magnetic field
dependence of satellite peak separation indicates excitation of
magnetoplasmons, as opposed to single-particle transitions for lower
concentrations\cite{finkelstein96,chapman98}.

\subsection{Shakeup effects in highly excited nanostructures}

All the above studies of the optical spectra
evolution from few- to many-particle cases were carried out for
{\em infinite} systems, i.e., those with {\em continuous} excitation
spectrum (at zero field). In such systems, the shakeup processes play
increasingly important role, as indicated by the emergence of the power-law
behavior at higher electron concentrations. However, these studies provide
no insight into the structure of many body states at the onset of
transition from discrete to continuous spectrum. In {\em finite-size} systems
with {\em discrete} excitation spectrum, the oscillator strengths of satellite
peaks characterize the amplitudes of corresponding shakeup transitions
which, in turn, are determined by the interaction matrix elements as well as
by statistical weights of the contributing many-body processes. With increasing
system size (at constant carrier concentration) new shakeup satellites should
emerge due to an increase in the number of many-body states available for
non-radiative transitions. In fact, the positions and magnitudes of shakeup
satellites represent the {\em fingerprint} of the system many-body excitation
spectrum. With further increase in the system size, the satellite peaks should
eventually merge; the energy dependence of their peaks {\em envelope} should
then follow the power-law lineshape of the continuous spectrum for the
corresponding infinite system.

\begin{figure}
\centering
\includegraphics[width=3in]{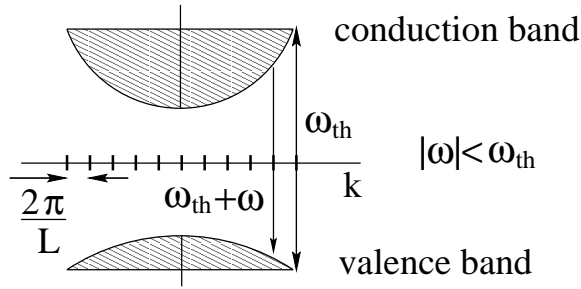}
\caption{\label{fig:e-h-FS} In highly-excited quantum-confined
  structures, photoexcited carriers form electron and
  hole Fermi seas in conduction and valence bands, respectively;
  $\omega_{th}$ is the energy distance between the corresponding Fermi levels.
  Recombination of electron-hole pairs, belonging to space-quantization
  energy levels, gives rise to discrete lines in the luminescence spectrum.
 }
\end{figure}

In the following, we consider in some detail the transition from
discrete to continuum spectra for the case of luminescence from highly
excited semiconductor nanostructures. We will restrict ourselves to
undoped semiconductors so that all carriers in conduction and valence
band are optically excited. The luminescence is preceded by a fast
carrier relaxation\cite{Toda}, so the recombination takes place when
the electron and hole gases are in their respective ground states. In
quantum wells, luminescence from high-density optically created
electron-hole gases was studied in
Refs. \cite{butov91-plasma,butov92-plasma,glasberg01}.  In confined
structures, such as quantum dots, electrons and holes fill
size-quantization energy states up to their respective Fermi levels in conduction
and valence bands (see Fig. \ref{fig:e-h-FS}). We will only consider
the higher-frequency domain of the emission spectrum corresponding to
frequencies not too far from the electron and hole Fermi edge's
separation.  For a noninteracting system, the emission lines would
correspond to the transitions between size-quantization levels in
conduction and valence bands which obey the selection rules, so the
the general expression (\ref{general}) becomes
\begin{eqnarray}
\label{non-inter}
I(\omega)\propto
\sum_n C_n\delta \Bigl[\omega+(\Delta_1+\Delta_2) n\Bigr],
\end{eqnarray}
where $\Delta_1$ and $\Delta_2$ are level spacings for electrons and
holes; $C_n$ are the oscillator strength which depend on $n$ only
{\em weakly} ($\omega<0$ is measured from the Fermi edge).

As discussed above, the many-body transitions due to the interactions
between carriers change qualitatively the form of the spectrum.
Namely, a removal of an electron-hole pair shakes up the respective Fermi seas
by causing them to emit Fermi sea excitations. Since in a {\em finite-size}
system, the energies of excitations are {\em quantized}, such a shakeup
would lead to the following spectrum,
\begin{equation}
\label{modified}
I(\omega)\propto\sum_{mn} C_{mn}
\delta \Bigl(\omega+m\tilde{\Delta}_1+n\tilde{\Delta}_2 \Bigr),
\end{equation}
rather than Eq.\ (\ref{non-inter}). Here $\tilde{\Delta}_1$ and
$\tilde{\Delta}_2$ are the level spacings renormalized by interactions.
The emission lines with $m\neq n$ are, thus, shakeup satellites.
All the information about many-body correlations in the system is encoded in
the oscillator strengths $C_{mn}$. In fact, $C_{mn}$, being governed by
interactions,  are {\em strong} functions of $m$ and $n$.

In general, the evaluation of coefficients $C_{mn}$ presents a major
challenge. Remarkably, in the case of 1D systems, $C_{mn}$ can be
calculated {\em analytically} when the number of carriers, $N$, is
large, but the emission spectrum is still
discrete\cite{shahbazyan01}. Such systems have recently been
manufactured and will be reviewed in the next section.  In this case,
the luminescence from 1D electron-hole system can be described within
the finite-size Luttinger liquid formalism\cite{haldane81}. Note that
the Luttinger liquid model was employed earlier for calculations of
the Fermi-edge optical properties of {\em infinite} 1D systems (with
and without defects) in
Refs. \cite{gogolin93,prokof'ev94,kane94,sassetti98,kramer00}.

\subsection{Semiconductor nanorings}

Properties of electron systems confined to a ring
have been a subject of a lot of studies during the
last decade (see Ref. \cite{chapelier95} for review).
Conceptually, the distinctive feature of the ring
geometry is that it is nonsimply connected. As a result,
the  orbital magnetism of electrons on a ring depends
periodically on the magnetic flux, $\Phi$, threading the ring.
Also, as a consequence of nonsilmply connectedness, the
many-electron ground state on a ring becomes chiral even
in a weak magnetic field, when $\Phi \ll \Phi_0$, where
$\Phi_0 = hc/e$ is the flux quantum. Nontrivial
magnetic and transport properties of electrons
in the ring geometry become observable when the ring
is small enough, so that the electron coherence length
exceeds the perimeter. This condition is met at low enough
temperatures. On the other hand, since the sensitivity
to the flux originates from the ring geometry, it persists
even if the electron elastic mean free path is smaller
than the perimeter, so that the overall character of the
electron transport is diffusive. The actual sizes of such
rings, that were studied  experimentally\cite{chapelier95},
were $\sim 10^3$ nm and more, so that the discreteness of
the quantum states in these zero-dimensional objects could
not be resolved.

Recently, a new technique for fabrication of the rings has
been introduced\cite{garcia97,lorke98,lorke99,pettersson00}.
In contrast to lithography\cite{chapelier95}, this
technique is based on the phenomenon of  self-assembly.
The rings are formed in two steps. The first step is the
conventional epitaxial growth of the array of  narrow gap (InAs)
quantum dots on the surface of the  wide-gap (GaAs)
substrate. Epitaxial islands are formed spontaneously in course
of this growth in order to minimize the elastic energy,
caused by the $7\%$  mismatch of the lattice constants of
InAs and GaAs. These islands are then made into dots
by covering them with another GaAs layer. The shape of
the dots  is lens (or pyramid)-like with $\sim 20$ nm in diameter
and $\sim 7$ nm in height. The second step is conversion of
the dots into volcano-shape
rings\cite{garcia97,lorke98,lorke99,pettersson00}, which is
achieved by annealing at the growth temperature.
The rings have the height of $\sim 2$ nm and the outer diameter
between 60 nm and 140 nm \cite{lorke00}. The center hole of
$\sim 20$ nm diameter is responsible for  nonsimply connectedness
of the confining potential for electrons and holes.

To demonstrate that this topology indeed dramatically changes the
response of electronic states in the rings to the magnetic flux, two
complementary spectroscopic techniques, capacitance-voltage
spectroscopy and far-infrared spectroscopy, were employed in
Ref. \cite{lorke00}. First technique measures the magnetic-field
dispersion of the ground state energy, whereas the second technique
provides information about the magnetic-field dependence of the
excitations. The measurements \cite{lorke00,petroff} have
revealed a cusps in charging energy and in positions of the minima in
far-infrared transmission as a function of magnetic field at $\Phi
\approx \Phi_0$. These cusps were identified with the change in the
angular momentum of the ground state. This conclusion was supported by
numerical calculations of Ref. \cite{emperador00}, which
reproduce the evolution of the peak positions in the far-infrared
absorption with magnetic field.

Experimental findings of Ref. \cite{lorke00} have triggered
theoretical studies of the single-electron states in quantum rings
\cite{barticevic02,monozon03}. In particular, the effect of
external electric filed\cite{barticevic02}
and impurity states in the ring geometry\cite{monozon03}
were addressed. Most interesting, however, are the many-body
effects in the ring geometry.

First experimental study of the many-body effects in quantum rings was
carried out in Ref. \cite{warburton00}, where the optical emission
from a charge-tunable ring was measured.  Similar to \cite{lorke00}
the tunability of the number of electrons on the ring was achieved due
to the presence of a gate electrode separated from a self-assembled
ring by a tunnel barrier. Changing the gate voltage allowed to add
electrons to the ring one-by-one.  Addition of each new electron
manifested itself as a step in the capacitance-voltage
characteristics.  The prime observation of Ref. \cite{warburton00} is
that all lines in the emission spectrum from the ring change abruptly
upon addition of an extra electron. Photoluminescence in
Ref. \cite{warburton00} was measured in the weak-excitation limit,
which corresponds to the ``classical'' shakeup situation (a single hole
plus degenerate electron gas).  Rearrangement of the entire emission
spectrum with addition of a single electron is a clear manifestation of
the many-body character of the luminescence from a ring.

Lithographically fabricated\cite{chapelier95} and self-assembled rings
differ by more than an order of magnitude in diameter. Characteristic
level spacing in self-assembled rings is rather large, $\sim 10$ meV.
This fact, and the finite bandgap offsets at GaAs/InAs boundaries,
restricts the maximal number of photoexcited electron-hole pairs as
well as the number of electrons, injected from the gate electrode into
the dot, to $\sim 10$. Taking into account the spin degeneracy, the
description of such a limited number of carriers in terms of a Fermi
sea is hardly adequate.  Therefore, in Ref. \cite{warburton00} the
language used to interpret the many-body emission spectrum was not a
shakeup (as in infinite system), but rather the electron-hole
recombination in the presence of ``spectators''.

Photoluminescence from quantum rings in external magnetic field is an
issue of conceptual interest for the following reason.  Photoexcited
electron and hole form an exciton, which is a neutral entity. Neutral
particle does not accumulate the Aharonov-Bohm phase in magnetic
field. Therefore, it might seem that, in contrast to the infrared
absorption, photoluminescence from the ring should not exhibit
oscillations with period $\Phi=\Phi_0$. This is, however, not the
case. The reason is that the exciton is a composite object.
Therefore, even bound electron and hole can tunnel in the opposite
directions and ``meet'' each other on the ``opposite'' side of the
ring. This process gives rise to the flux sensitivity of the exciton.
Theoretical studies of magneto-optical properties of the neutral and
charged quantum ring excitons were reported in
Refs.
\cite{chaplik95,govorov97,roemer00,roemer'00,hu00,hu01,roemer01,meier01,ulloa01,hu'01,govorov02}.
Moreover, the mechanism\cite{chaplik95,roemer00} of the flux
sensitivity of the exciton in a ring geometry was extended in
Refs. \cite{vedernikov01,chaplik02} to other neutral excitations
(plasmons).

In the experimental paper Ref. \cite{haft02}, a weak anomaly in
the luminescence spectrum of a neutral exciton in a self-assembled
ring was interpreted as a possible manifestation of the Aharonov-Bohm
effect.

\section{LUTTINGER LIQUID THEORY OF LUMINESCENCE FROM HIGHLY
  EXCITED NANORINGS}


\subsection{General expression for emission rate}

\begin{figure}
\centering
\includegraphics[width=4in]{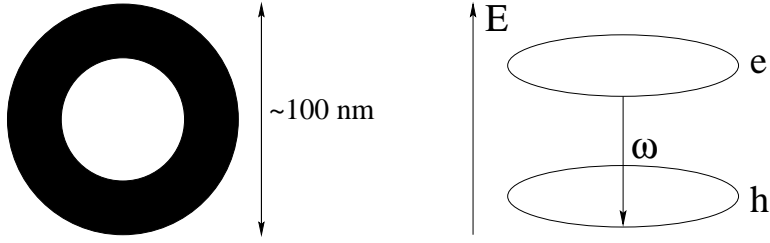}
\caption{\label{fig:LL-ring} Schematic representation of an
 electron-hole pair recombination in highly-excited Luttinger liquid ring}
\end{figure}

Here we outline the general formulas for recombination of an
electron-hole pair belonging to Luttinger liquid rings in conduction
and valence bands. (see Fig. \ref{fig:LL-ring}).
We start with the two-component Luttinger liquid model on a
ring\cite{shahbazyan97} with Hamiltonian $H_1+H_2+H_{int}$, where
$H_j$ describe noninteracting electrons ($j=1$) and holes ($j=2$) with
linearized dispersions (the slopes are determined by the Fermi
velocities $v_j$); $H_{int}$ describes the interactions between
carriers via screened potential $U(x)$.  The electron-hole
recombination rate is given by the Golden rule\cite{mahan} [compare
Eq. (\ref{general})]
\begin{equation}
\label{golden}
W(\omega)=\frac{2\pi}{L}\sum_f|\langle f |T|i\rangle|^2
\delta(E_i-E_f-\omega)
=\frac{1}{L}\int_{-\infty}^{\infty}dt
e^{-i\omega t}\langle i|T^{\dag}(t)T(0)|i\rangle,
\end{equation}
where $E_i$ and $E_f$ are the energies of initial (ground) and final
(with electron-hole pair removed) states, and
\begin{equation}
\label{tran}
T=T_{+}+T_{-},
\mbox{\hspace{5mm}}
T_{\pm}=d\int_{0}^{L}dx \psi_{2\mp}(x)\psi_{1\pm}(x)
\end{equation}
is the dipole transition operator. Here $\psi_{i\pm}$ are annihilation
operators for left ($-$) and right ($+$) moving carriers, $d$ is the
interband dipole matrix element, and $L$ is the ring
circumference. Note that recombination occurs between left (right)
electrons and right (left) holes.  The recombination rate is then
expressed via a four-particle Green function,
\begin{equation}
\label{prob-D}
W(\omega)=
d^2\int_{0}^{L}dx\int_{-\infty}^{\infty}dt
e^{-i\omega t}\bigl[ D_{+}(x,t)+D_{-}(x,t)\bigr]
=
d^2\bigl[ D_{+}(\omega)+D_{-}(\omega)\bigr],
\end{equation}
where
\begin{equation}
\label{D}
D_{\pm}(x,t)=
\langle
\psi_{2\mp}^{\dag}(x,t)\psi_{1\pm}^{\dag}(x,t)
\psi_{1\pm}(0)\psi_{2\mp}(0)\rangle.
\end{equation}

\subsection{Bosonisation of electron-hole Hamiltonian}

In order to evaluate $D_{\pm}(x,t)$ for a two-component Luttinger
liquid\cite{larkin74,matveev93,penc93}, we use the bosonization
technique on a ring\cite{haldane81,shahbazyan01,shahbazyan97}.
The right/left fermion fields are presented as
\begin{equation}
\label{fer}
\psi_{j\alpha}(x)
=L^{-1/2}:e^{i\varphi_{j\alpha}(x)}:
=(2\pi \epsilon)^{-1/2}e^{i\varphi_{j\alpha}(x)+i\alpha \pi x/L},
\end{equation}
where $\epsilon$ is a cutoff, while right/left ($\alpha=\pm$) bosonic
fields $\varphi_{j\alpha}(x)$ are related to the corresponding
densities as
\begin{equation}
\label{dens}
\rho_{j\alpha}(x)=\frac{\alpha}{2\pi}
\frac{\partial \varphi_{j\alpha}(x)}{\partial x}.
\end{equation}
The bosonic field has a decomposition
\begin{equation}
\label{bos-free}
\varphi_{j\alpha}(x)=\varphi_{j\alpha}^0+\alpha N_{j\alpha}2\pi x/L
+\bar{\varphi}_{j\alpha}(x),
\end{equation}
where the number operator $N_{j\alpha}$ and its conjugate $\varphi_{j\alpha}^0$
satisfy the commutation relations
\begin{equation}
\label{zero-comm}
[N_{j\alpha},\varphi_{k\beta}^0]=i\delta_{kj}\delta_{\alpha\beta},
\end{equation}
and the periodic field
$\bar{\varphi}_{j\alpha}(x)=\bar{\varphi}_{j\alpha}(x+L)$ has a standard form,
\begin{equation}
\label{bos-free-per}
\bar{\varphi}_{j\alpha}(x)=
\sum_{k}\theta(k\alpha)\sqrt{\frac{2\pi}{L|k|}}e^{-|k|\epsilon/2}
\biggl( e^{ikx}a_{kj} + e^{-ikx}a_{kj}^{\dag}\biggr),
\end{equation}
with $a_{kj}$ and $a_{kj}^{\dag}$ satisfying boson commutation relations
[$\theta(x)$ is the step function]. The boundary condition for the fermion
fields, $\psi_{j\alpha}(x+L)=(-1)^{N_j}\psi_{j\alpha}(x)$,
depends on the parity of the number of particles,
$N_j=2N_{j\alpha}$
($N_{j\alpha}$ may have half-integer eigenvalues),
and follows from the factorization
\begin{equation}
\label{fer-factor}
\psi_{j\alpha}(x)=L^{-1/2}:e^{\bar{\varphi}_{j\alpha}(x)}:
e^{\varphi_{j\alpha}^0}e^{\alpha N_{j\alpha}2\pi i x/L}.
\end{equation}
The Hamiltonian $H=H_0+H_{int}$ is quadratic in boson fields:
\begin{equation}
\label{H-free}
H_0=\sum_{j\alpha}\frac{v_j}{4\pi}
\int_0^Ldx
\Biggl[\frac{\partial \varphi_{j\alpha}(x)}{\partial x}\Biggr]^2,
\end{equation}
and
\begin{equation}
\label{H-int}
H_{int}=\frac{1}{2}\sum_{jl}
\int_0^Ldx\int_0^Ldy
\Biggl[\sum_{\alpha}\frac{\alpha}{2\pi}
\frac{\partial \varphi_{j\alpha}(x)}{\partial x}\Biggr]
U_{jl}(x-y)
\Biggl[\sum_{\beta}\frac{\beta}{2\pi}
\frac{\partial \varphi_{l\beta}(y)}{\partial y}\Biggr],
\end{equation}
where $U_{jl}(x)$ is the screened potential. Using
Eqs.\ (\ref{bos-free}) and (\ref{bos-free-per}), the
zero-modes can be separated from the bosonic part. The total
Hamiltonian is then just a
sum, $H=H_0+\bar{H}$, of the zero-modes contribution,
\begin{equation}
\label{H-zero}
H_0=\frac{\pi}{L}\sum_{jl\alpha\beta}
N_{j\alpha}
\biggl(v_j\delta_{jl}\delta_{\alpha\beta}+\frac{u_{jl}}{2}\biggl)
N_{l\beta},
\end{equation}
and periodic bosonic fields contribution
\begin{equation}
\label{H-per}
\bar{H}=
\sum_{qjl} e^{-|q|\epsilon}|q|
\biggl[v_j\delta_{jl}a_{qj}^{\dag}a_{qj}
+\frac{u_{jl}}{4}(a_{qj}^{\dag}+a_{-qj})(a_{-ql}^{\dag}+a_{ql})
\biggr],
\end{equation}
where $u_{jl}=\pi^{-1}\int dx U_{jl}(x)$.

In order to calculate the correlation functions, the Hamiltonian $\bar{H}$
has to be brought to
the canonical form. This is done in two steps. First, we perform a
two-component Bogolubov transformation in order to
eliminate the cross-terms with opposite momenta,
\begin{equation}
\label{bogol}
a_{qj}=\sum_l\bigl(X_{jl}b_{ql}+Y_{jl}b_{-ql}^{\dag}\bigr),
~~~
\sum_l\bigl(X_{jl}X_{ln}^{\dag}-Y_{jl}Y_{ln}^{\dag}\bigr)=\delta_{jn}.
\end{equation}
We then obtain
\begin{eqnarray}
\label{H-bar1}
\bar{H}=\sum_{qjl} e^{-|q|\epsilon}|q|
b_{qj}^{\dag}\bigl(X^{\dag}-Y^{\dag}\bigl)_{jl}v_l
\bigl(X-Y\bigl)_{ln}b_{qn},
\end{eqnarray}
where the matrices $X$ and $Y$ must satisfy
\begin{eqnarray}
\label{bogol-eq}
\sum_{lm}
\bigl(X^{\dag}+Y^{\dag}\bigl)_{jl}
(u_{lm}+v_l\delta_{lm})
\bigl(X+Y\bigl)_{mn}
=
\sum_l
\bigl(X^{\dag}-Y^{\dag}\bigl)_{jl}v_l
\bigl(X-Y\bigl)_{ln}.
\end{eqnarray}
Second, we diagonalize the Hamiltonian (\ref{H-bar1}) by first presenting the
matrices $X$ and $Y$ as
\begin{eqnarray}
\label{XY}
X=\cosh\lambda \,O,
~~~
Y=\sinh\lambda\, O,
\end{eqnarray}
where $\lambda_{jl}=\lambda_j\delta_{jl}$ is the diagonal matrix of Bogolubov
angles $\lambda_j$ and $O$ is an orthogonal matrix, and then by introducing new
boson operators
\begin{eqnarray}
\label{bos-new}
c_{qj}=\sum_lO_{jl}b_{ql}.
\end{eqnarray}
The Hamiltonian $\bar{H}$ then takes the form
\begin{eqnarray}
\label{H-diag}
\bar{H}=\sum_{qj} e^{-|q|\epsilon}|q| \tilde{v}_jc_{qj}^{\dag}c_{qj},
\end{eqnarray}
with renormalized Fermi velocities
\begin{eqnarray}
\label{vel-renorm}
\tilde{v}_j=e^{-2\lambda_j}v_j.
\end{eqnarray}
Note that old and new boson operators are related  as
\begin{eqnarray}
\label{old-new}
a_{qj}=\cosh\lambda_j\, c_{qj}+ \sinh\lambda_j\, c_{-qj}^{\dag}.
\end{eqnarray}
Using the decomposition (\ref{XY}), Eq.\ (\ref{bogol-eq}) takes the
form $\tilde{O}AO=0$, where $\tilde{O}$ is the transposed matrix,
and the matrix $A$ is given by
\begin{eqnarray}
\label{bogol-matr}
A_{jl}=u_{jl}e^{\lambda_j+\lambda_l}+
\delta_{jl}v_j\bigl(e^{2\lambda_j}-e^{-2\lambda_j}\bigr).
\end{eqnarray}
The Bogolubov angles $\lambda_j$ are found from the condition
that all the eigenvalues of $A_{jl}$ vanish.
In a two-component case, this yields
\begin{eqnarray}
\label{two-comp}
&&
e^{-2\lambda_1}=
\sqrt{Q
\frac{v_1+u_{11}-v_2 Q}
{v_1 Q-v_2-u_{22}}},
\mbox{\hspace{5mm}}
e^{-2\lambda_2}=Q /e^{-2\lambda_1},
\nonumber\\ &&
\mbox{\hspace{5mm}}
Q=\sqrt{\biggl(1+\frac{u_{11}}{v_1}\biggr)
\biggl(1+\frac{u_{22}}{v_2}\biggr)
-\frac{u_{12}^2}{v_1v_2}}.
\end{eqnarray}
Correspondingly, the level spacings are now
$\tilde{\Delta}_j=2\pi \tilde{v}_j/L$.

\subsection{Calculation of the Green function}

With the Hamiltonian (\ref{H-diag}), the time-dependence of new operators is
standard,
\begin{eqnarray}
\label{bos-time}
c_{kj}(t)=e^{-i\tilde{v}_j|k|t}c_{kj}.
\end{eqnarray}
Using the relation (\ref{old-new}), the periodic fields take the form
\begin{equation}
\label{bos-int-per}
\bar{\varphi}_{j\alpha}(x,t)=
\sum_{k}\sqrt{\frac{2\pi}{L|k|}}e^{-|k|\epsilon/2}
\Bigl[\theta(k\alpha)\cosh\lambda_j+
\theta(-k\alpha)\sinh\lambda_j\Bigr]
\Bigl(e^{ikx-i\tilde{v}_j|k|t}c_{kj}
+e^{-ikx+i\tilde{v}_j|k|t}c_{kj}^{\dag}\Bigr).
\end{equation}
The time-dependence of zero-modes can be easily obtained using
equation of motion,
\begin{equation}
\label{zero-eom}
i\partial \varphi_{j\alpha}^0/\partial t=[\varphi_{j\alpha}^0,H_0],
\end{equation}
with the zero-mode Hamiltonian (\ref{H-zero}). The final expression for the total
time-dependent bosonic field reads
\begin{equation}
\label{bos-int}
\varphi_{j\alpha}(x,t)=\varphi_{j\alpha}^0
+\alpha N_{j\alpha}2\pi (x-\alpha v_j t)/L
-\sum_{l\beta}u_{jl}N_{l\beta}\,\pi t/L
+\bar{\varphi}_{j\alpha}(x,t).
\end{equation}
We are now in position to calculate the Green functions. For this,
we separate out annihilation and creation parts of the periodic field
(\ref{bos-int-per}),
\begin{equation}
\label{bos-per-decomp}
\bar{\varphi}_{j\alpha}(x,t)=\bar{\varphi}_{j\alpha}^{-}(x,t)
+\bar{\varphi}_{j\alpha}^{+}(x,t),
\end{equation}
which satisfy the following commutation
relations
\begin{equation}
\label{bos-int-per-comm}
[\bar{\varphi}_{j\alpha}^{-}(x,t),\bar{\varphi}_{j\alpha}^{+}(x',t')]
=
\ln f_{\alpha}(z_{j\alpha}-z'_{j\alpha})
+\mu_i\ln \Bigl[
f_{\alpha}(z_{j\alpha}-z'_{j\alpha}) f_{-\alpha}(z_{j,-\alpha}-z'_{j,-\alpha})
\Bigr],
\end{equation}
with
\begin{equation}
\label{z}
z_{j\alpha}=x-\alpha\tilde{v}_jt.
\end{equation}
Then we present the fermion operator (\ref{fer}) in the normal-ordered form,
\begin{eqnarray}
\label{fer-normal}
&&
\psi_{j\alpha}(x,t)
=\psi_{j\alpha}^{0}(x,t)\bar{\psi}_{j\alpha}(x,t),
\nonumber\\
&&
\psi_{j\alpha}^{0}(x,t)=
e^{iv_j(1+u_{jj}/2)\pi t/L}e^{i\varphi_{j\alpha}^0}
e^{i\alpha N_{j\alpha}2\pi z_{j\alpha}/L
-i\sum_{l\beta}u_{jl}N_{l\beta}\,\pi t/L},
\nonumber\\
&&
\bar{\psi}_{j\alpha}(x,t)=
L^{-1/2}\Bigl(2\pi\epsilon/L\Bigr)^{\mu_j}
e^{i\bar{\varphi}_{j\alpha}^{+}(x,t)}
e^{i\bar{\varphi}_{j\alpha}^{-}(x,t)},
\end{eqnarray}
where we again separated out zero-mode and periodic parts.
Using Eq.\ (\ref{fer-normal}) together with commutators
(\ref{zero-comm}) and (\ref{bos-int-per-comm}), the correlation functions can
be straightforwardly calculated. For the Green function (\ref{D}), we
obtain
\begin{equation}
\label{D-bos-app}
D_{\alpha}(x,t)=
\biggl(\frac{2\pi\epsilon}{L}\biggr)^{2(\mu_2+\mu_2)}
\frac{e^{-it\delta_P-it\delta_u}}{L^2}
\Bigl[f_{\alpha}(z_{1\alpha})\Bigr]^{1+\mu_1}
\Bigl[f_{-\alpha}(z_{1,-\alpha})\Bigr]^{\mu_1}
\Bigl[f_{-\alpha}(z_{2,-\alpha})\Bigr]^{1+\mu_2}
\Bigl[f_{\alpha}(z_{2\alpha})\Bigr]^{\mu_2},
\end{equation}
where
\begin{equation}
\label{mu}
\mu_i=\sinh^2\lambda_i
\end{equation}
is the Luttinger liquid interaction parameter, while $\delta_P=\pi
(v_1+v_2)/L$ and $\delta_u=\pi (u_{11}+u_{22}+2u_{12})/2$ are energy shifts
due the changes in the parity of electron and hole numbers and in the Coulomb
energy, caused by a removal of an electron-hole pair. The coordinate
dependence of $D_{\alpha}(x,t)$ is determined by (with $\alpha=\pm$)
\begin{eqnarray}
\label{I-bos}
f_{\alpha}(z_{j\alpha})
=
\frac{1}{1-e^{i\alpha (2\pi z_{j\alpha}/L +\alpha i\bar{\epsilon})}},
\end{eqnarray}
where $\bar{\epsilon}=2\pi\epsilon/L$ is the dimensionless cutoff. We assume
that the screened interaction is the same for electrons and holes,
$u_{11}=u_{22}=-u_{12}=u$, so that $\delta_u=0$. After neglecting the parity
phase $\delta_P$ (which can be absorbed into the frequency),
we finally write
\begin{eqnarray}
\label{D-bos}
D_{\pm}(x,t)=
\frac{\bar{\epsilon}^{2(\mu_2+\mu_2)}}{L^2}\,
\Bigl[f_{\pm}(z_{1\pm})\Bigr]^{1+\mu_1}
\Bigl[f_{\mp}(z_{1\mp})\Bigr]^{\mu_1}
\Bigl[f_{\mp}(z_{2\mp})\Bigr]^{1+\mu_2}
\Bigl[f_{\pm}(z_{2\pm})\Bigr]^{\mu_2}.
\end{eqnarray}
The interaction strength is characterized by the ratio
$u/v_j$, where
\begin{eqnarray}
\label{u}
u= \frac{1}{\pi} \int dx U(x)
\end{eqnarray}
is the Fourier of screened potential; this ratio represents the average
(screened) interaction in units of the (bare) level spacing near the Fermi
energy. For weak interactions, $u/v_j\ll 1$, we have
\begin{eqnarray}
\label{weak}
\mu_j
\simeq \left(\frac{u}{4 v_j}\right)^2,
\mbox{\hspace{5mm}}
\tilde{\Delta}_j\simeq \Delta_j\left(1+\frac{u}{2v_j}\right).
\end{eqnarray}
Note finally that the above calculation is easily generalized if the ring is
penetrated by a magnetic flux $\phi$. In this case, the electron and hole
number operators should be shifted by flux-dependent constants,
$N_{1\alpha}\rightarrow N_{1\alpha}+\alpha\phi/\phi_0$ and
$N_{2\alpha}\rightarrow N_{2\alpha}-\alpha\phi/\phi_0$,
where $\phi_0$ is the flux quantum. This results in the replacement
$\delta_P\rightarrow \delta_P(1-2\alpha\phi/\phi_0)$ in Eq.\ (\ref{D-bos-app}).

\subsection{Derivation of oscillator strengths}

The  correlator $D_{\alpha}(x,t)$ is periodic in
variables $z_{j\alpha}$. In order to
carry out the integration in Eq.\ (\ref{prob-D})
we first perform the Fourier expansion of
$\bigl[f_{\alpha}(z_{j\alpha})\bigr]^{\nu}$ as
\begin{eqnarray}
\label{I-fourier}
\Bigl[f_{\alpha}(z_{j\alpha})\Bigr]^{\nu}=
\sum_n
b_{\nu}(n)e^{i\alpha 2\pi nz_{j\alpha}/L},
~~~
b_{\nu}(n)
=
\frac{\sin \pi\nu}{\pi}
B(n+{\nu},1-{\nu}),
\end{eqnarray}
Substituting this Fourier expansion into Eq.\ (\ref{D-bos}),
the Green function $D_{\alpha}(\omega)$ takes the form
\begin{eqnarray}
\label{D-moment}
D_{\pm}(\omega)=
\bar{\epsilon}^{2(\mu_1+\mu_2)}
\sum_{\{n\}}
b_{1+\mu_1}(n_1)b_{\mu_1}(n'_1)b_{1+\mu_2}(n_2)b_{\mu_2}(n'_2)
\Lambda_{\pm}(\omega,\{n\}),
\end{eqnarray}
with
\begin{eqnarray}
\label{Lambda}
\Lambda_{\pm}(\omega,\{n\})=
\frac{1}{L^2}\int dt \int_0^{L} dx
\exp\biggl[
-i\omega t \pm
i\frac{2\pi}{L}(n_1z_{1\pm} -n'_1z_{1\mp}
-n_2z_{2\mp}
+n'_2z_{2\pm})
\biggr]
\nonumber\\
=\frac{2\pi}{L}
\delta_{n_1-n'_1,n_2-n'_2}
\delta\biggl[\omega+\frac{2\pi \tilde{v}_1}{L}(n_1+n'_1)
+\frac{2\pi \tilde{v}_2}{L}(n_2+n'_2)
\biggr],
\end{eqnarray}
where we absorbed the parity shift $\delta_P$ into $\omega$.
The Kronecker delta and the delta-function reflect the conservation of
momentum and energy, respectively. Thus, we obtain
\begin{eqnarray}
\label{D-final}
D_{\alpha}(\omega)=\frac{2\pi}{L}\sum_{mn}C_{mn}
\delta\Bigl(\omega +\tilde{\Delta}_1m+\tilde{\Delta}_2n\Bigr),
\end{eqnarray}
with
\begin{equation}
\label{Cmn}
C_{mn}=
\bar{\epsilon}^{2(\mu_1+\mu_2)}
\sum_{l}
b_{1+\mu_1}[(m+n)/2-l]b_{1+\mu_2}(n-l)b_{\mu_1}[(m-n)/2+l]b_{\mu_2}(l).
\end{equation}
From Eq.\ (\ref{D-final}), the emission spectrum (\ref{modified}) follows.
Finally, using integral representation for the Beta-function in
Eq.\ (\ref{I-fourier}) we arrive at
\begin{equation}
 \label{Cmn-int}
 C_{mn}=
 \int_{-\pi}^{\pi}
 \frac{d\phi_1d\phi_2d\phi_3}{(2\pi)^3}
 \frac{\bar{\epsilon}^{2(\mu_1+\mu_2)}\,
 e^{-\frac{i}{2}(\phi_1+\phi_2)(m+n)+\frac{i}{2}\phi_3(m-n)}}
 {\Bigl(1-e^{i\phi_1}\Bigr)^{1+\mu_1}
 \Bigl(1-e^{i\phi_2}\Bigr)^{1+\mu_2}
 \Bigl(1-e^{i(\phi_2-\phi_3)}\Bigr)^{\mu_1}
 \Bigl(1-e^{i(\phi_1+\phi_3)}\Bigr)^{\mu_2}}.
\end{equation}
Note, that the sum in Eq.\ (\ref{modified}) is constrained by the
selection rule that $m$ and $n$ are of the same parity, i.e., the
combinations
\begin{equation}
N=(m+n)/2,
~~~
M=(m-n)/2,
\end{equation}
which enter into the rhs of Eq.\ (\ref{Cmn-int}), are integers, as can
be seen from Eq.\ (\ref{Lambda}). This is a result of the linear
dispersion of electrons and holes near the Fermi levels.

It is easy to see that Eq.\ (\ref{Cmn-int}) correctly reproduces
the non-interacting limit. Indeed, upon setting $\mu_i=0$, the
integral over $\phi_3$ yields $C_{mn}=\delta_{mn}$.
Another important limiting case $m,n\gg 1$  corresponds to the
transitions well away from the Fermi edge. In this case, the main
contribution to the integral (\ref{Cmn-int}) comes from the domain
$\phi_1+\phi_2\sim (m+n)^{-1}\ll 1$. Within this domain, one can
neglect the difference between $\phi_1$ and $-\phi_2$ in the last two
factors in the denominator. Then the integrals over $\phi_1,\phi_2$
factorize, yielding
\begin{eqnarray}
\label{Cmn-est}
C_{mn}=
\frac{\Gamma(N+1+\mu_1)\Gamma(N+1+\mu_2)}
{\Gamma(1+\mu_1)\Gamma(1+\mu_2)[\Gamma(N+1)]^2}\, K(M),
\end{eqnarray}
with
\begin{equation}
\label{Kn}
K(M)=
\int_{-\pi}^{\pi}\frac{d\phi}{2\pi}
\frac{\bar{\epsilon}^{2(\mu_1+\mu_2)}\, e^{iM\phi}}
{\Bigl(1-e^{-i\phi}\Bigr)^{\mu_1}
\Bigl(1-e^{i\phi}\Bigr)^{\mu_2}}
=\frac{\epsilon^{2(\mu_1+\mu_2)}(-1)^M\Gamma(1-\mu_1-\mu_2)}
{\Gamma(1-M-\mu_1)\Gamma(1+M-\mu_2)},
\end{equation}
where $\Gamma(x)$ is the Gamma-function.
It can be seen from Eq.\ (\ref{Kn}) that, for a given $N$, the
oscillator strengths, $C_{mn}$, fall off as
$C_{mn}\propto |M|^{\mu_1+\mu_2-1}$ with increasing
$|M|=\frac{1}{2}\vert m-n \vert$.
This slow power-law decay reveals strong correlations within
electron-hole system on a ring. Finally, using the large $x$
asymptotics of $\Gamma(x)$, we obtain the expression for the
oscillator strengths valid for $|m-n|\gg 1$,
\begin{equation}
\label{asym}
C_{mn}=\frac{\bar{\epsilon}^{2\mu}\Gamma(1-\mu)}
{\Gamma(1+\mu_1)\Gamma(1+\mu_2)}\frac{\sin\pi\tilde\mu}{\pi}
\left(\frac{m+n}{2}\right)^{\mu}\left|\frac{m-n}{2}\right|^{\mu-1},
\end{equation}
where
\begin{equation}
\label{mus}
\mu=\mu_1+\mu_2,
~~~~
\tilde \mu =\frac{1}{2}\mu+
\frac{1}{2}(\mu_1-\mu_2){\rm sgn} (m-n).
\end{equation}


\section{\large FINE STRUCTURE OF THE EMISSION SPECTRUM}

The general expression (\ref{Cmn-int}) determines the heights of the
emission peaks, while the {\em order} of the peaks with
different $\{m,n\}$ is governed by the $\delta$-functions in
Eq.\ (\ref{modified}), which ensure the energy conservation.
Therefore, this order depends crucially on the relation
between $\tilde{\Delta}_1$ and $\tilde{\Delta}_2$. Moreover,
a {\em commensurability} between $\tilde{\Delta}_1$ and
$\tilde{\Delta}_2$ leads to accidental degeneracies in
the positions of the emission lines.
However, in order to establish the general properties
of the spectrum, it is instructive to consider first
several cases of commensurate $\tilde{\Delta}_1$ and
$\tilde{\Delta}_2$.

\subsection{Symmetric case}

We start with the symmetric case $\tilde{\Delta}_i=\tilde{\Delta}/2$
(and, hence, $\mu_i=\mu/2$).  The peak positions, as determined by
Eq.\ (\ref{modified}), coincide with those for single-particle
transitions, $|\omega|=N\tilde{\Delta}$. The corresponding oscillator
strengths can be straightforwardly evaluated from Eq.\ (\ref{Cmn-int})
as
\begin{eqnarray}
\label{cN-exact}
c_N=\sum_{M}C_{N+M,N-M}=\Biggl[\int_{-\pi}^{\pi}\frac{d\phi}{2\pi}
\frac{\bar{\epsilon}^{\mu}\, e^{-iN\phi}}
{\bigl(1-e^{i\phi}\bigr)^{1+2\mu}}\Biggr]^2.
\end{eqnarray}
For $N\gg 1$, the denominator of the integrand  can be expanded, yielding
\begin{eqnarray}
\label{cN}
c_N
\simeq (\bar{\epsilon} N)^{2\mu}
=\biggl|\frac{\bar{\epsilon}\omega}{\tilde{\Delta}}\biggr|^{2\mu}.
\end{eqnarray}
Note that single-particle oscillator strengths correspond to $c_N=1$.
We thus conclude that interactions affect strongly the peak heights
for $|\omega/\tilde{\Delta}|^{2\mu}\gg 1$, i.e., in the high frequency
domain.  In fact, even for an arbitrary relation between
$\tilde\Delta_1$ and $\tilde\Delta_2$, the crossover between
``single-particle'' and ``many-body'' domains of the spectrum is
governed by the dimensionless parameter
\begin{equation}
\label{xi}
\xi=\mu\ln\frac{|\omega|}{\tilde{\Delta}_1+\tilde{\Delta}_2}.
\end{equation}

\subsection{Commensurate case}

Now consider the case when the level spacings in the conduction and
valence band are commensurate:
\begin{equation}
\frac{\tilde{\Delta}_1}{\tilde{\Delta}_2}=\frac{p}{q},
\end{equation}
where $p$ and $q$ are integers. Introducing a notation
\begin{equation}
\label{Delta}
\tilde{\Delta}=\tilde{\Delta}_1+\tilde{\Delta}_2,
\end{equation}
the Green function (\ref{D-final}) takes the form
\begin{eqnarray}
\label{D-rat-gen}
D_{\alpha}(\omega)
=\frac{2\pi}{L}\sum_{mn}C_{mn}
\delta\biggl(\omega +\tilde{\Delta}\frac{mp+nq}{p+q}\biggr)
=\frac{2\pi}{L}\sum_k C_k\delta(\omega+\tilde{\Delta} k/Q),
\end{eqnarray}
where $Q=p+q$, $P=p-q$,
and
\begin{eqnarray}
\label{Ck-gen}
C_k
=\sum_{mn}\delta_{k,mp+nq}C_{mn}
=\sum_{MN}\delta_{k,MP+NQ}C_{N+M,N-M}
\nonumber\\
=\sum_{MN}\delta_{k-MP,NQ}
C_{\frac{k}{Q}+M\bigl(1-\frac{P}{Q}\bigr),
\frac{k}{Q}-M\bigl(1+\frac{P}{Q}\bigr)}.
\end{eqnarray}
Using the relation
\begin{eqnarray}
\sum_N\delta_{k,NQ}=\frac{1}{Q}\sum_{l=0}^{Q-1}e^{-i2\pi l k/Q},
\end{eqnarray}
the oscillator strengths
can be presented as
\begin{eqnarray}
\label{Ck-final-gen}
C_k=\frac{1}{Q}\sum_{l=0}^{Q-1} e^{-i2\pi l k/Q}f_l(k),
\end{eqnarray}
with
\begin{eqnarray}
\label{f-genl}
f_l(k)=\sum_Me^{i2\pi l M P/Q}
C_{\frac{k}{Q}+M\bigl(1-\frac{P}{Q}\bigr),
\frac{k}{Q}-M\bigl(1+\frac{P}{Q}\bigr)}.
\end{eqnarray}
Using integral representation (\ref{Cmn-int}), the sum over $M$ can be
explicitly performed. For $k/Q=|\omega|/\tilde{\Delta}\gg 1$, the
resulting expression for coefficients $f_l$ takes the form
\begin{eqnarray}
\label{fl-int-asympt-gen}
f_l(k)=
\int_{-\infty}^{\infty}
\frac{d\phi_1d\phi_2}{(2\pi)^2}
\frac{\bar{\epsilon}^{2(\mu_1+\mu_2)}\,
e^{-i(\phi_1+\phi_2)k/Q}}
{(-i\phi_1)^{1+\mu_1}(-i\phi_2)^{1+\mu_2}
\Bigl(1-s_l-is_l[\phi_2+(\phi_1+\phi_2)P/Q]\Bigr)^{\mu_1}}
\nonumber\\
\times
\frac{1}
{\Bigl(1-s_l^{\ast}-is_l^{\ast}[\phi_1-(\phi_1+\phi_2)P/Q]\Bigr)^{\mu_2}},
\end{eqnarray}
where $s_l=e^{i2\pi lP/Q}$. The $l$-dependence of $f_l(k)$ is
determined by the relative magnitude of $Q/k$ and $|1-s_l|$:
\begin{eqnarray}
\label{fl-est1}
f_l(k)
\simeq
\biggl|\frac{\bar{\epsilon} k}{Q}\biggr|^{2(\mu_1+\mu_2)}
\end{eqnarray}
for $k/Q\ll |1-s_l|^{-1}$, and
\begin{eqnarray}
\label{fl-est2}
f_l(k)\simeq
\Biggl(\frac{\bar{\epsilon}^2 k/Q}{1-s_l}\Biggr)^{\mu_1}
\Biggl(\frac{\bar{\epsilon}^2 k/Q}{1-s_l^{\ast}}\Biggr)^{\mu_2}
\end{eqnarray}
for  $k/Q\gg |1-s_l|^{-1}$,
with the two estimates matching at $k/Q\sim |1-s_l|^{-1}$.

Summarising, each "parent" single-particle peak, corresponding to $k=nQ$ in
Eq. (\ref{D-rat-gen}), acquires $Q-1$ shakeup satellites separated in energy by
$\tilde{\Delta}/Q$. Such equidistant distribution of satellite positions is
due to periodicity in the excitation spectrum caused by commensurate level
spacings in conduction and valence band. The oscillator strengths depends
strongly on frequency, $\omega/\tilde{\Delta}=k/Q$, as indicated by
Eqs. (\ref{Ck-final-gen},\ref{fl-est1},\ref{fl-est2}). The emergence of
satellite peaks with increasing $\omega/\tilde{\Delta}$ is demonstated
below for the simplest case when each single-particle peak acquires
just one satellite.

\subsection{Case $\tilde\Delta_1=3\tilde\Delta_2$}

Consider now the case $\tilde\Delta_1=3\tilde\Delta_2$
(and thus $\mu_2\simeq 9\mu_1$) which corresponds to the doubling of
luminescence peaks number as compared to noninteracting case.
Indeed, as follows from  Eq.\ (\ref{modified}), here the peak spectral
positions are given by
\begin{eqnarray}
\label{l/2}
\frac{|\omega|}{\tilde{\Delta}}=\frac{n}{2}.
\end{eqnarray}
For $P=2$ and $Q=4$, so that $s_l=(-1)^l$, the coefficients $f_l$
take two different values depending on the parity of $l$ [see
Eqs. (\ref{fl-est1},\ref{fl-est2})],
\begin{eqnarray}
\label{fl-even}
&&
f_{even}(k)
\simeq
\biggl|\frac{\bar{\epsilon} k}{4}\biggr|^{2\mu}
=\biggl|\frac{\bar{\epsilon} \omega}{\tilde{\Delta}}\biggr|^{2\mu},
\\
\label{fl-odd}
&&
f_{odd}(k)\simeq
\biggl|\frac{\bar{\epsilon}^2 k}{8}\biggr|^{\mu}
=\biggl|\frac{\bar{\epsilon}^2 \omega}{2\tilde{\Delta}}\biggr|^{\mu},
\end{eqnarray}
yielding [see Eq. (\ref{Ck-final-gen})],
\begin{eqnarray}
\label{Ck-even-odd}
C_k=
\biggl|\frac{\bar{\epsilon} \omega}{\tilde{\Delta}}\biggr|^{2\mu}
\frac{1+(-1)^k}{2}
\frac{1+e^{i\pi k/2}\bigl|\frac{\tilde{\Delta}}{2\omega}\bigr|^{\mu}}{2},
\end{eqnarray}
with $\mu=\mu_1+\mu_2$. Obviously, $C_k=0$ for $k$ odd. For $k$ even, we have
\begin{eqnarray}
\label{Ck-even-even}
&&
C_{4l}\simeq
\biggl|\frac{\bar{\epsilon} \omega}{\tilde{\Delta}}\biggr|^{2\mu}
\frac{1+\bigl|\frac{\tilde{\Delta}}{2\omega}\bigr|^{\mu}}{2},
\\
&&
C_{4l+2}\simeq
\biggl|\frac{\bar{\epsilon} \omega}{\tilde{\Delta}}\biggr|^{2\mu}
\frac{1-\bigl|\frac{\tilde{\Delta}}{2\omega}\bigr|^{\mu}}{2},
\end{eqnarray}
and we finally obtain
\begin{equation}
\label{D=2}
D_{\pm}(\omega)
=\frac{2\pi}{L}
\biggl|\frac{\bar{\epsilon}\,\omega}{\tilde{\Delta}}\biggr|^{2\mu}
\sum_l\Biggl[
\frac{1+\bigl|\frac{\tilde{\Delta}}{2\omega}\bigr|^{\mu}}{2}
\,\delta(\omega+\tilde{\Delta} l)
+
\frac{1-\bigl|\frac{\tilde{\Delta}}{2\omega}\bigr|^{\mu}}{2}
\,\delta\bigg[\omega+\tilde{\Delta}\bigg(l+\frac{1}{2}\bigg)\bigg]
\Biggr],
\end{equation}
with $\mu=\mu_1+\mu_2\simeq 10\mu_1$ and
$\tilde{\Delta}=\tilde{\Delta}_1+\tilde{\Delta}_2=4\tilde\Delta_1$.

\begin{figure}
\centering
\includegraphics[width=4in]{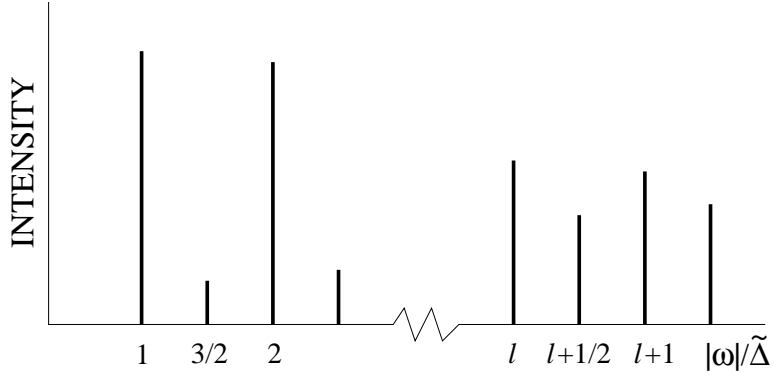}
\caption{\label{fig:1-sat} Emission spectrum for
  $\tilde\Delta_1=3\tilde\Delta_2$. In low-frequency domain, the
  single-particle peaks acquire weak many-body shakeup
satellites. In high-frequency domain, the heights of the
parent ($|\omega|/\tilde{\Delta}=l$) and satellite
($|\omega|/\tilde\Delta=l+1/2$) peaks are close to each other.
 }
\end{figure}

The above result illustrates how the structure of the spectrum evolves
as the frequency departs from the Fermi edge.
For $\xi=\mu\ln\bigl|\frac{\omega}{\tilde{\Delta}}\bigr|\ll 1$, each
single-particle peak, $|\omega|=l\tilde\Delta$
acquires a weak shakeup satellite at $|\omega|=(l+\frac{1}{2})\tilde\Delta$.
In the opposite limit, $\xi \gg 1$,
the oscillator strength of an ``integer'' peak is equally
redistributed
between the doublet components. The crossover frequency,
 separating the ``single-particle'' and the developed many-body
domains of the spectrum is determined by the condition
$\xi\sim 1$, or, $\omega\sim \tilde{\Delta}e^{1/\mu}$.
The spectrum (\ref{D=2})
is schematically depicted in Fig.\ \ref{fig:1-sat}.

\subsection{General structure of the smission spectrum}

Let us turn to the structure of the spectrum in the general case of
incommensurate $\tilde \Delta_1$ and $\tilde \Delta_2$.  We start from
the observation that the peak positions can be classified by
``generations''.  Namely, once a peak $\{m,0\}$ (or $\{0,n\}$) emerges
at $\omega=\omega_m=-m\tilde\Delta_1$ (or
$\omega=\omega_n=-n\tilde\Delta_2$), it is followed by next
generations of peaks $\omega_m^{(k)}=\omega_m-k(\tilde\Delta_1+
\tilde\Delta_2)$ or $\omega_n^{(k)}=\omega_n-k(\tilde\Delta_1+
\tilde\Delta_2)$ repeating with a period
$\tilde\Delta=\tilde\Delta_1+\tilde\Delta_2$. Thus, for a crude
description of the spectrum away from the Fermi edge it is convenient
to divide the frequency region $\omega<0$ into the intervals of width
$\tilde\Delta$.

The number of peaks within the spectral interval
$\{-\vert\omega\vert,-\vert\omega\vert-\tilde\Delta\}$ is the number
of integers satisfying the conditions
\begin{equation}
\label{number-condition}
\vert\omega\vert < m\tilde\Delta_1+
n\tilde\Delta_2<|\omega|+\tilde\Delta.
\end{equation}
This number is equal to
\begin{equation}
\label{number}
{\cal N}_{\omega}=
\frac{|\omega|\tilde\Delta}{2\tilde\Delta_1\tilde\Delta_2},
\end{equation}
where we assumed $\vert\omega\vert\gg \tilde\Delta$ and took into account the
parity restriction. From Eq.\ (\ref{number}) we
find the peak density
\begin{equation}
\label{density}
g_{\omega}=
\frac{{\cal N}_{\omega}}{\tilde\Delta}
=\frac{\vert\omega\vert}{2\tilde\Delta_1\tilde\Delta_2}.
\end{equation}
It also follows from Eq. ({\ref{number})
that
\begin{equation}
\label{dN}
\delta {\cal N}={\cal N}_{\omega-\tilde\Delta}-{\cal N}_{\omega}
=\frac{\tilde\Delta^2}{2\tilde\Delta_1\tilde\Delta_2}
\end{equation}
generations start within each interval independently of frequency.
Since the heights of consecutive peaks within the
interval $\tilde\Delta$ vary non-monotonically, it is natural to
characterize these heights by the distribution function
\begin{equation}
\label{dist}
F({\cal C})=
\frac{1}{2g_{\omega}}
\!
\int_0^{\infty}dm\, dn
\,\delta\bigl(\omega+m\tilde\Delta_1+n\tilde\Delta_2\bigr)\,
\delta\bigl(C_{mn}-{\cal C}\bigr),
\end{equation}
where $C_{mn}$ is given by Eq.\ (\ref{asym}).
Here we made use of the fact that ${\cal N}_{\omega}\gg 1$ by treating
$m$ and $n$ as continuous variables. The prefactor in Eq.\ (\ref{dist})
ensures the normalization
$\left(\int_0^{\infty}d{\cal C} F({\cal C})=1\right)$. It is easy to
see that $F({\cal C})$ is nonzero in the interval
(${\cal C}_{min},{\cal C}_{max}$), where (hereafter we omit the cutoff)

\begin{eqnarray}
{\cal C}_{min}=
{\rm min}\bigl\{2\mu_{1,2}
\bigl|\frac{\tilde\Delta_{1,2}}{\omega}\bigr|^{1-2\mu}\bigr\},
~~~~~
{\cal C}_{max}=
2\bigl|\frac{\omega}{\tilde\Delta}\bigr|^{\mu}{\rm max}\{\mu_{1,2}\}.
\end{eqnarray}
Within this wide interval, the distribution function falls off as
\begin{equation}
F({\cal C})\sim \left(\frac{{\cal C}_0}{{\cal C}}\right)^{2+\mu},
~~~
{\cal C}_0=
\mu\Biggl|\frac{\omega}{4}\Biggl(\frac{1}{\tilde{\Delta}_1}
+\frac{1}{\tilde{\Delta}_2} \Biggr)\Biggr|^{2\mu-1}
\end{equation}
where ${\cal C}_0$ is the {\em typical} value of the oscillator strength. On
the other hand, the {\em average} oscillator strength, which can be easily
calculated from  Eq.\ (\ref{dist}), is equal to
$\overline{\cal C}=\mu^{-1}{\cal C}_0\gg {\cal C}_0$.
The distribution function $F({\cal C})$ is schematically depicted in
Fig.\ \ref{fig:general}.  The fact that $\overline{\cal C}$ {\em decreases} with
$|\omega|$ can be understood in the following way. As it is seen from
Eq.\ (\ref{cN}), in the symmetric case, with only a single
peak per interval $\tilde\Delta$, the peak heights increase with
$|\omega|$ as $|\frac{\omega}{\tilde\Delta}|^{2\mu}$.
In the general case,
this spectral intensity gets redistributed between ${\cal N}_{\omega}$
different peaks. Thus,
%
%
\begin{equation}
\overline{\cal C}
\sim {\cal N}_{\omega}^{-1}\biggl|\frac{\omega}{\tilde\Delta}\biggr|^{2\mu}
\propto |\omega|^{2\mu-1}.
\end{equation}
\begin{figure}
\centering
\includegraphics[width=4in]{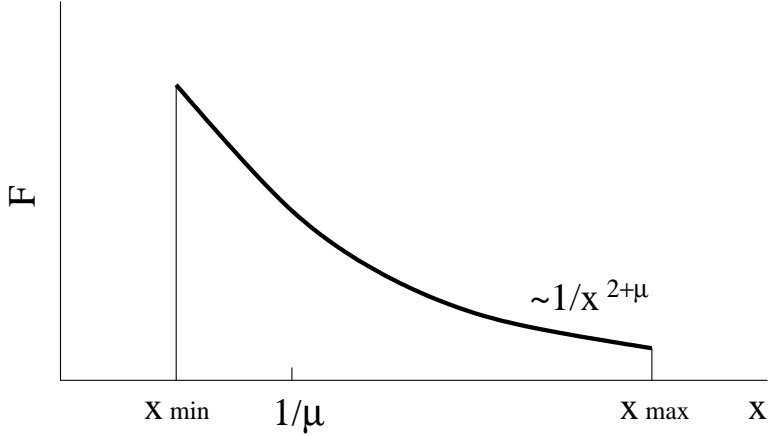}
\caption{\label{fig:general} The distribution function (\ref{dist}) of
  the peak heights within the interval $\tilde\Delta$ is plotted
  schematically versus $x={\cal C}/{\cal C_0}$. The minimal value of
  $x$ is $x_{min}\sim 1$, while $x_{max}\sim |\omega/\tilde\Delta
  |^{1-\mu}\gg 1$. The point $x=\mu^{-1}$ corresponds to the average
  oscillator strength.
}
\end{figure}

\section{\large CONCLUDING REMARKS}


The underlying origin of the multitude of many-body lines in the emission spectrum
from quantum-confined systems is that a recombination is accompanied by shakeup
processes whose number increases as the frequency deviates from the Fermi edge.  The
theoretical value of the dimensionless interaction parameter $\mu$ is determined by
the ratio of screened interaction $U$ to the level spacings $\tilde\Delta_1$ and
$\tilde\Delta_2$ at the corresponding Fermi levels.  Both quantities depend on the
number of excited carriers, $N$, which in turn is determined by the excitation
intensity. This, and the sensitivity of the screening to the details of experimental
setup, lead to a common ambiguity in the theoretical determination of $\mu$. For
example, in quantum wires, the value of $\mu$ measured in resonant tunneling
experiments\cite{yacoby00a,yacoby00b}, was significantly larger then theoretical
estimates.  Concerning the estimates for $\tilde\Delta_1$ and $\tilde\Delta_2$, in the
experimental paper Ref.\ \cite{warburton00} on luminescence from ring-shape dots the
total energy separation $\tilde\Delta$ between the lowest levels was approximately 5
meV.  This value comes almost exclusively from the conduction band, due to the large
ratio of the electron and hole effective masses.  Both $\tilde\Delta_1$ and
$\tilde\Delta_2$ increase linearly with increasing $N$. This implies that the shake-up
processes within the hole system are experimentally much more relevant than those for
electrons.

T.V.S. was supported by the Army High Performance Computing Research Center under the
auspices of the Department of the Army, Army Research Laboratory under Cooperative
Agreement No DAAD19-01-2-0014, and by the National Science Foundation under Grant Nos
DMR-0304036 and DMR-0305557. M.E.R. was supported by the National Science Foundation
under Grant No. DMR-0202790 and by Petroleum Research Fund under Grant No. 37890-AC6.

\vspace{10mm}
%


%

\end{document}